\documentstyle{article}

\begin{document}

{\bf ON THE INTERPRETATION OF RELATIVISTIC SPIN NETWORKS AND THE
BALANCED STATE SUM}

\bigskip

by Louis Crane, Mathematics department, KSU

\bigskip

{\bf 1. THE GEOMETRIC INTERPRETATION}

\bigskip

In [1], a state sum model was proposed which had a natural
relationship to the quantum theory of general relativity in 3+1 dimensions.
In order to establish an actual connection to general relativity,
however, it will be necessary to show that the stationary phase terms
in the sum can be interpreted as discrete approximations to solutions
of the vacuum Einstein equations.

The purpose of this note is to show how terms in the state sum may be
interpreted as geometries, ie as discretized metrics on the underlying
triangulated manifold. We still do not have a proof that steepest
descent corresponds to Einstein, but a plausible argument can be made.

The geometric interpretation we propose makes use of some operators
introduced by Barbieri in [2] for tetrahedra in three dimensions, then
extended to relativistic or 4d tetrahedra in [3]. See also the
discusion in [4]. 

The operator Barbieri constructs is a quantum mechanical analog of the
determinant which computed the volume of a tetrahedron. Classically if
$ V_i, i=1,2,3; V_i=(x_i, y_i, z_i) $ are the three vectors connecting
one vertex of a tetrahedron in $R^3$ (thought of as posessing a
natural inner product) to the other 3 vertices, then the
volume of the tetrahedron is given by

\bigskip

$V=1/6 det[x_i, y_i, z_i]$ . (1)

\bigskip

As Barbieri points out, this volume can instead be computed from the
determinant

\bigskip

$det[V_2 \times V_3, V_3 \times V_1, V_1 \times V_2]$ ; (2) 

\bigskip

which is exactly the matrix of cofactors of the first determinant, and
consequently has determinant equal to its square. The second
determinant is formed from the bivectors associated to three of the
faces of the tetrahedron. Since we are in 3 dimensions, bivectors can
be naturally interpreted as vectors. This is, of course, the
mathematical meaning of the cross product of vectors. If we make any
other choice of three of the four faces of a tetrahedron, the
determinant is the same.

The picture both in [1] and [2] is to regard the bivectors on
faces of a tetrahedron as fundamental variables rather than the
lengths or directions of edges, to reinterpret the bivectors as
angular momenta, and then to quantize them by replacing them with
representations of su(2) as is usual in the quantization of the spin. 

Thus it is important to replace an expression for the volume from
edges by one from the bivectors on the faces. Barbieri then writes an
operator for the quantum theory which is the quantum analog of
determinant [2],which he denotes {\bf U} :

\bigskip

{\bf $U=J_1 . J_2 \times J_3 $}

\bigskip

which is the obvious quantum analog of determinant (2).

In [3]. Barbieri writes a similar expression to his {\bf U} operator for a
four dimensional quantum tetrahedron as defined in [1], ie for a
four-valent vertex for a relativistic spin net. 

For a tetrahedron in $R^4$, however, a scalar volume does not contain
all possible information. The volume form of a tetrahedron in $R^4$,
obtained analogously to our determinant (1) by wedging together the
three 1-forms dual to the three edge vectors connecting any vertex to
the other three; is a
3-form whose magnitude gives the volume of the tetrahedron (times 6)
and whose ``direction'' gives the orientation of the hyperplane
containing the tetrahedron. This 3-form has four components, which
can be expressed as determinants analogous to four copies of
determinant (1). They are the determinants of the four $3 \times 3$
minors of the  $4 \times 3$ matrix formed by the 3 4-vectors.

Using definitions analogous to Barbieri's, it is possible to assign to
a ``relativistic quantum tetrahedron'' ie to a tetravalent vertex for
a relativistic spin net, a vector of four operators, analogous to the
four components of the 3-form in the classical situation. The
definition is straightforward. One begins with the expressions for the
two copies of su(2) in the splitting of so(4) or so(3,1):

\bigskip

$J^x_{ \pm} = J^{yz} \pm J^{xt}$

\bigskip

or respectively

$J^x_{ \pm } = J^{yz} \pm iJ^{xt}$ etc.

\bigskip

The i in these formulas is the only thing to distinguish the Euclidean
and Minkowski signature cases in this approach to quantum gravity.

We can simply invert these and obtain our six bivector operators for
the spin on each face of a quantum tetrahedron. We can then form our
four 4d analogs of the { \bf U } operator as $3 \times 3$ matrices of
operators. Let us call them { \bf $ U_x, U_y, U_z, U_t$ } or { \bf $
U^{ \rightarrow} $ }.

For example, the $dy \wedge dz \wedge dt $ component of the 3-form
associated to the tetrahedron generated by the edges numbered 1,2,3 is
given by the determinant of the $ 3 \times 3$ matrix of operators
whose ith row is $ J^i_{yz}, J^i_{zt}, J^i_{tx} $.

Now let us see what we have in the case of a ``quantum 4-simplex'', ie
a 15J symbol in the category of relativistic spin nets. This is what
we get for one labelling of one 4-simplex in the state sum of [1].
For each tetrahedron in the boundary of the labelled 4-simplex, we now
have a 4-vector of operators { \bf $ U_x, U_y, U_z, U_t$ } as above. 

It is now possible to form invariant combinations of these vector operators,
namely their inner products with themselves and each other.
This gives us the quantum volumes of the tetrahedra and their
hyperdihedral angles. These suffice to specify a  geometry
corresponding to definite values of the U operator products.

(The reader  familiar with [1] will doubtless note the similarity
between the proof of the classical geometrical theorem there and this
construction).

It is very important for us that we now have operators corresponding
to the hyperdihedral angles, since in a four dimensional triangulated
geometry it is the sums of the hyperdihedral angles around faces which
specify the curvature.

We close with conjectures about how the quantum dihedral angles we
have defined will behave when we combine 15J symbols into the state
sum in [1].

{ \bf Conjecture 1;} The evaluation of relativistic 15j symbol has a
rapid oscillation of phase with respect to the variation of a quantum
hypergeometric angle as defined above.

{ \bf Conjecture 2:} If the state sum of [1] is considered without the
balance constraints, stationary phase will impose that the sum of the
hyperdihedral angles around a face be $2 \pi$ in the classical limit;
the effect of the constraints is to reduce the stationary phase
condition to a combination of sums of hyperdihedral angles converging
to Ricci flatness in the classical limit.

The verification of these will require a good deal of work. Research
on them is under way.

\bigskip

{\bf 2. ON THE `` ULTRAVIOLET'' LIMIT}

\bigskip

The suggestion was made in [1] that the states to be considered in the
quantum theory of gravity would be asymptotically self dual, so that the
result of the balanced state sum at some particular triangulation
would be a good approximation to an exact result. We want to make a
more physical version of this proposal here. 

Let us make the hypothesis that the universe began in a self dual
state. This is plausible because the boundary conditions at the event
horizon of a black hole are self dual [5,6]. This would mean that the
state of the whole universe remained self dual. The state of the whole
universe, however, is not what an observer experiences. The
wavefunction of the universe has gone through many collapses or
bifurcations, depending on whether one likes the Copenhagen or many
worlds picture. In any event, the history of the universe has in
effect measured a very large number of areas, giving rise to non self
dual bivectors in many places. However, if one takes a triangulation T
fine enough to contain everything that any macroscopic physical
subsystem of the universe has measured, then the states in any
tetrahedron strictly finer than the triangulation would be self
dual. Since the self dual states are in effect governed by a
topological state sum, it is unnecessary to make a finer
triangulation, and the state sum on T will give an exact answer.

It appears that this interpretation of the model may lead to testable
consequences, in the form of history dependence of some physical
property of the universe. This requires further thought.

\bigskip

{\bf CONCLUSIONS}

\bigskip

It is, of course, much to soon to speak of any real success for this
model. What can be fairly said is that many deep issues of quantum
gravity, ranging from the ultraviolet behavior to the Minkowskian
signature difficulties, can be attacked within it by well defined
computations. Given the difficult history of quantum gravity, however,
this is at least noteworthy.

\bigskip

{\bf BIBLIOGRAPHY}

\bigskip

1.J. W. Barrett and L. Crane Relativistic Spin Networks and Quantum
gravity gr-qc 9709028

\bigskip

2. A. Barbieri Quantum Tetrahedra and Simplicial Spin Networks, gr-qc 9707010

\bigskip

3. A.Barbieri Space of Vertices of Relativistic Spin Networks gr-qc 9709076

\bigskip

4. J. Baez, Spin Foam models gr-qc 9709052

\bigskip

5. L. Smolin, J. Math. Phys. 36 6417 (1995).

\bigskip

6. A. Ashtekar, J. Baez, A. Corichi and K. Krasnov, Quantum Geometry
and black Hole Entropy gr-qc 9710007

\end{document}